# Comparison of the generalized centroid with Gaussian and quadratic peak localization methods

**SHANNON CAMPBELL**

*Micro Encoder Inc.*
*shannonc@microen.com*

**We examine a new method for peak localization, the centroid of the data raised to some power, which we call the generalized centroid. We derive the peak localization uncertainty for the generalized centroid and compare it with the Cramer-Rao lower bound for both Gaussian and quadratic fits (with Gaussian signal and noise). We find that the centroid of squares and the Gaussian fit yield the best results in both one and two dimensions. We perform similar analysis with a Lorentz-like signal and find that the centroid of cubes and the nonlinear least squares fit provide the best results. We support our derivations with simulations, and also show simulation results when the maximum function is used to initially estimate the peak location.**

## Introduction

There are many applications that require finding the center of a peak in astronomy, biology, spectroscopy, and optics [1-8]. Gaussian fitting (assuming a nonlinear least-squares method) is common and the peak uncertainty is well known [9-13]. The centroid method has been studied in detail, e.g. [1,7], surprisingly however, the centroid of squares, centroid of cubes, and centroid of data raised to higher powers do not seem to have been examined. We refer to this as the generalized centroid and analyze it now. The quadratic fit is also commonly used, but we have not found a detailed analysis of it and derive the Cramer-Rao minimum bound here. To our knowledge, there has been no direct analytical comparison of these aforementioned methods, although numerical comparisons of some of these methods have been performed in [5].

We compare peak localization uncertainties for a Gaussian signal with Gaussian noise for the generalized centroid, Gaussian fit, and quadratic fit methods through both standard error analysis and numerical simulations. The uncertainties of all methods have the same basic dependence on the signal parameters of amplitude, $A$, width, $w$, signal noise, $\sigma_s$, and sample interval, $\Delta x$. Specifically, the variance of the peak location is given by

$$\sigma^2_{\text{peak loc.}} = C(k) \frac{w \Delta x \sigma_s^2}{A^2}. \tag{1}$$

The coefficient of proportionality, $C(k)$, changes depending on the method, and is a function of the range, $k$, over which data is acquired.

For large ranges, defined here as more than three times width of the curve, the centroid of squares, which is a computationally inexpensive method, and Gaussian fitting have the same and the smallest variance (of the methods we examine). The centroid and the quadratic uncertainties become increasingly large as the range increases. The centroid of cubes and the centroid of data raised to higher powers approach finite values larger than that of the centroid of squares.

We also examine the generalized centroid and the quadratic fit when the signal has a Lorentz-like shape with Gaussian noise. Instead of using the Gaussian fit with this signal, we find the Cramer-Rao minimum bound of the variance for the peak location assuming a nonlinear least-squares method that fits the data using a Lorentz-like equation. We find that the centroid of cubes and the non-linear least squares fit give the same and the smallest variance (of the methods we examine) for large ranges.

In many implementations, peak localization algorithms first find a local maximum and then use some range, or window, of data centered about that local maximum. The location of this maximum is affected by noise, but we cannot mathematically analyze this first algorithmic step. We present simulation results with and without the maximum value included and find that generally, the effects of finding the maximum value become less significant as the range increases. We will discuss these results in detail.

The paper is organized as follows. We first derive the peak location uncertainties for a Gaussian signal using the generalized centroid, quadratic fit, and non-linear least squares fit in one and two dimensions. We then derive peak localization uncertainties for these same methods when the signal has a Lorentz-like shape but only for one dimension. We then compare our analysis with simulations that both contain and omit the algorithmic step of first finding the maximum value. We finish by summarizing our results and briefly discussing some directions for future work.

# Peak localization uncertainties for Gaussian signal with Gaussian noise

## Generalized centroid

We define the generalized centroid, $c_p$, as

$$c_p = \frac{\sum_{i=-m}^{m} x_i s_i^p}{\sum_{i=-m}^{m} s_i^p}, \quad (2)$$

where the $x_i$ are the signal locations and the $s_i$ represent the measured signal at those locations. There are $N = 2m + 1$ data points acquired at some interval $\Delta x$, and that interval must be smaller than the curve width for the signal to be appropriately sampled. For $p = 1$ we call it the centroid and for $p = 2$ we refer to it as the centroid of squares, etc. The measured signal has an associated Gaussian noise, $\sigma_s$, that is constant. To reduce visual clutter, we will ignore the indices on the summation symbols.

We propagate error using the variance formula [14, 15]. Taking the first derivative of the generalized centroid with respect to the $n^{th}$ signal value yields

$$\frac{\partial c_p}{\partial s_n} = \frac{p s_n^{p-1} x_n}{\sum s_i^p} - \frac{p s_n^{p-1} \sum x_i s_i^p}{(\sum s_i^p)^2}. \quad (3)$$

The last term on the r.h.s. contains the definition of the centroid and because $c_p \approx 0$ we ignore it. We calculate the variance by summing over all $N$ derivatives in quadrature,

$$\sigma_{c_p}^2 = \sigma_s^2 \sum \left(\frac{\partial c_p}{\partial s_i}\right)^2 = \sigma_s^2 \frac{p^2 \sum s_i^{2p-2} x_i^2}{(\sum s_i^p)^2}. \quad (4)$$

We approximate the sums in eq. (4) with integrals and obtain

$$p^2 \sum s_i^{2p-2} x_i^2 \approx \frac{p^2}{\Delta x} \int_{-kw}^{kw} dx\, x^2 \left(A e^{-\frac{x^2}{2w^2}}\right)^{2p-2} \quad (5a)$$

$$\approx \frac{A^{2p-2} p^2 w^3}{2\Delta x (p-1)} \left(\frac{\sqrt{\pi}\, \Phi(k(p-1))}{\sqrt{p-1}} - 2k e^{-k^2(p-1)}\right). \quad (5b)$$

$\Phi$ is the error function. We integrate from $-kw$ to $+kw$, and later, when describing the range, refer to $k$ only. Also, we integrate over a symmetric range. Had we used the maximum function to find the peak value and define the center of the range, the asymmetry in the range would make analysis more complex.

The denominator is approximated by

$$\left(\sum s_i^p\right)^2 \approx \left(\frac{1}{\Delta x} \int_{-kw}^{kw} dx\, \left(A e^{-x^2/2w^2}\right)^p\right)^2 \quad (6a)$$

$$\approx \frac{A^{2p} w^2}{\Delta x^2} \left(\frac{2\pi}{p}\right) \Phi\left(k\sqrt{\frac{p}{2}}\right)^2. \quad (6b)$$

Combining the numerator and denominator yields the coefficient of proportionality for the generalized centroid

$$C_{Gc}(k,p) = \frac{p^3 \left(\frac{\sqrt{\pi}\, \Phi(k(p-1))}{\sqrt{p-1}} - 2k e^{-k^2(p-1)}\right)}{4\pi(p-1)\, \Phi(k\sqrt{p/2})^2}. \quad (7)$$

For the case of the centroid, $p = 1$, the coefficient is

$$C_{Gc}(k,1) = \frac{k^3}{3\pi\, \Phi(k/\sqrt{2})^2}. \quad (8)$$

The centroid integrates over the noise in an additive fashion and the variance increases without bound as $k \to \infty$. For $p > 1$, the noise is suppressed by multiplication with the signal, which decays to zero rapidly and the variance approaches a finite value as $k \to \infty$.

For $k \to \infty$ and $p > 1$ the coefficient is

$$C_{Gc}(p) = \frac{p^3}{4\sqrt{\pi}(p-1)^{3/2}}, \quad (9)$$

and this has a minimum value at $p = 2$.

For two dimensions, with an underlying signal of

$$S(x,y) = A e^{-\left(\frac{x^2}{2w_x^2} + \frac{y^2}{2w_y^2}\right)}, \quad (10)$$

we perform the same type of analysis and find that the variance in the $x$-direction is

$$C_{Gcx}(k,p) = \frac{p\, \Phi(k(p-1))}{2\pi^{1/2}(p-1)^{1/2}} \frac{C_{Gc}(k,p)}{\Phi(k\sqrt{p/2})^2} \quad (11)$$

multiplied by $w_x \Delta y \Delta x \sigma_s^2 / w_y A^2$. For $p = 1$, the coefficient is

$$C_{Gcx}(k,1) = \frac{k^4}{3\pi^2\, \Phi(k/\sqrt{2})^4}. \quad (12)$$

The variance in the $y$-direction is direction found by switching the $x$- and $y$- labels on the width parameters.

For $k \to \infty$ and $p > 1$ the two-dimensional generalized centroid has the following variance in the $x$-direction

$$C_{Gcx}(p) = \frac{p^4}{8\pi(p-1)^2}, \quad (13)$$

and this has a minimum value at $p = 2$. Also, unlike the one-dimensional case, the variance at $k = 0$ is non-trivial, so we mention it here,

$$C_{Gcx}(0,p) = p^2/12. \quad (14)$$

## Quadratic fit

We find the Cramer-Rao minimum bound for the localization variance for the quadratic function by following the path laid out in [11]. The minimum parameter variances are bounded by the inverse of the Fisher information matrix, which is the expectation of the Hessian. Let us start with a quadratic equation

$$f(x, p_j) = p_1 - p_2(x - p_3)^2. \quad (15)$$

We write the quadratic equation in this form to make the peak location an explicit parameter.

We then create a cost function to be minimized,

$$\chi^2 = \sum \frac{[f(x_i, p_j) - s_i]^2}{\sigma_s^2}. \quad (16)$$

To form the Hessian, we take second order partial derivatives of the above function with respect to the parameters $p_j$. The Hessian is the matrix below summed over all data points

$$\begin{bmatrix} 1 & -(x_i - p_3)^2 & 2p_2(x_i - p_3) \\ -(x_i - p_3)^2 & (x_i - p_3)^4 & -2p_2(x_i - p_3)^3 \\ 2p_2(x_i - p_3) & -2p_2(x_i - p_3)^3 & 4p_2^2(x_i - p_3)^2 \end{bmatrix}. \quad (17)$$

Approximating the sum over the data points with an integration over the range -$kw$ to $kw$ yields,

$$\begin{bmatrix} 2kw & -\frac{2}{3}k^3w^3 & 0 \\ -\frac{2}{3}k^3w^3 & \frac{2}{5}k^5w^5 & 0 \\ 0 & 0 & \frac{8}{3}p_2^2k^3w^3 \end{bmatrix}. \quad (18)$$

To complete analysis, we need to solve for the value of $p_2$ that results from a least squares quadratic fit to a Gaussian curve as a function of $k$. This can be done and is

$$p_2 = \frac{15A}{8k^5w^2}\left((k^2-3)\sqrt{2\pi}\Phi(k/\sqrt{2}) + 6ke^{-k^2/2}\right). \quad (19)$$

The coefficient for the quadratic peak center variance is thus

$$C_{Gq}(k) = \frac{8k^7}{75\left((k^2-3)\sqrt{2\pi}\Phi(k/\sqrt{2}) + 6ke^{-k^2/2}\right)^2}. \quad (20)$$

The above equation diverges as $k \to 0$ and as $k \to \infty$.

For a two-dimensional signal as defined in eq. (10) we find that the variance for the quadratic fit in the x-direction is

$$C_{Gqx}(k) = \frac{4k^8}{75\left[\Phi\left(\frac{k}{\sqrt{2}}\right)\left(\pi(k^2-3)\Phi\left(\frac{k}{\sqrt{2}}\right) + 3\sqrt{2\pi}ke^{-\frac{k^2}{2}}\right)\right]^2}, \quad (21)$$

(multiplied by $w_x\Delta y\Delta x\sigma_s^2/w_yA^2$). This equation diverges as $k \to 0$ and as $k \to \infty$.

**Gaussian fit**

The uncertainties for the Gaussian fit to a Gaussian signal with Gaussian noise are well known [9-13]. However, the calculations in those papers were done as $k \to \infty$ because it is a practical approximation. We want to perform a comparison for finite $k$ because, in our experience, there are many applications in which the window size is tightly constrained.

We do not go through all the steps to derive the uncertainty associated with the Gaussian fit, because it is analogous to what we did for the quadratic fit and simply state that the coefficient for the Gaussian fit peak localization variance is

$$C_G(k) = \frac{2}{\sqrt{\pi}\Phi(k) - 2ke^{-k^2}}. \quad (22)$$

For a two-dimensional Gaussian signal, as defined in eq. (10), the coefficient for the variance in the x-direction is

$$C_{Gx}(k) = \frac{2}{\sqrt{\pi}\Phi(k)(\sqrt{\pi}\Phi(k) - 2ke^{-k^2})}. \quad (23)$$

It can be shown that the centroid of squares and the Gaussian peak localization variances have the same coefficient of proportionality as $k \to \infty$, namely, $2/\sqrt{\pi}$ in one dimension and $2/\pi$ in two dimensions. It is of course interesting and useful to know that the centroid of squares can yield results identical to the more computationally complex Gaussian fit (for large $k$).

## Peak localization uncertainties for Lorentz-like signal with Gaussian noise

It is not clear why the centroid of squares should be equivalent to Gaussian fitting in terms of peak localization variance for large $k$. Out of curiosity, we examine the peak localization uncertainties using a different signal, a Lorentz-like function defined as,

$$f(x, p_j) = \frac{p_1}{(p_2 + (x - p_3)^2)}. \quad (24)$$

For this function we define the width as $w = \sqrt{p_2}$ and the amplitude is $A = p_1/p_2$. All integrals are carried to $\pm kw$ as before. We calculate the variances for the different peak localization methods as described previously for the Gaussian signal.

The variance for the generalized centroid is complicated and long, as performed with Mathematica, and not revealing when written as a function of $k$. We do not transcribe it here. But we do display the variance as $k \to \infty$,

$$C_{Lc}(p) = \frac{\Gamma(p+1)^2\Gamma(2p-7/2)}{2\sqrt{\pi}\Gamma(p+1/2)^2\Gamma(2p-2)}, \quad (25)$$

where $\Gamma$ is the gamma function. The minimum value is $4/\pi$ at $p = 3$.

The peak localization coefficient for the quadratic fit is

$$C_{Lq}(k) = \frac{2k^7}{75(-3k + (3 + k^2)\tan^{-1}(k))^2}. \quad (26)$$

The above equation goes to infinity as $k \to 0$ and as $k \to \infty$.

The peak localization coefficient for a non-linear least-squares fit with the Lorentz-like function is

$$C_L(k) = \frac{6(1+k^2)^3}{(-3k + 8k^3 + 3k^5 + 3(1+k^2)^3\tan^{-1}(k))}. \quad (27)$$

Graphs of the above equations and comparisons with simulations are provided in the next section.

The centroid of cubes and the non-linear least-squares fit have the same coefficient of proportionality as $k \to \infty$, namely, $4/\pi$.

## Comparison with simulations

We now compare our theoretical analysis with simulation results. We perform two types of simulations, one that is consistent with the analysis above and a more realistic simulation that includes the maximum function to find the center of the range.

The simulations are carried out as follows; we generate a Gaussian (or a Lorentz-like) signal of random amplitude, width, and sample interval and then add Gaussian noise. The amplitude of the noise is always less than the amplitude of the signal by a factor of 20-200. The width ranges from 0.5 to 2.5 and the sample interval ranges from ½ to 1/50th of the width. The range is fixed at $6w$. We calculate the generalized centroid for $p = 1, 2, 3$ and 4. For one set of simulations the range is symmetrically centered about zero (the known center of the signal) and for the other set, the maximum value is the center.

For each set of randomly generated signal parameters, we create multiple samples with independent Gaussian noise to compute the variance. For the one-dimensional signals, we create 1600 sets of parameters and for each set, generate 200 samples. For the two-dimensional simulations, we make 600 parameter sets with 60 samples each. We calculate the peak location for multiple values of $k$ at intervals of $0.1w$.

For each parameter set, we calculate the variance of the peak locations. Then, using all parameter sets and their associated variances, we perform a single parameter linear least-squares fit to find the coefficient of proportionality, i.e.

$$\sigma_{sim}^2(k) \approx C_{sim}(k)\frac{w\Delta x\,\sigma_n^2}{A^2}. \quad (28)$$

In Figure 1, we display the uncertainty from simulation (labeled symbols) and compare with our analysis (curves) for a Gaussian signal with Gaussian noise without first finding the maximum. The simulations and theory are consistent with one another.

We show the uncertainty (the square root of the variance) in this and the following graphs because it allows for a more detailed visualization of the data. Also, estimated confidence intervals for each data point are similar in size to the symbols, so we do not display them. This is true for all data except that in Figure 6, where we do display some confidence intervals.

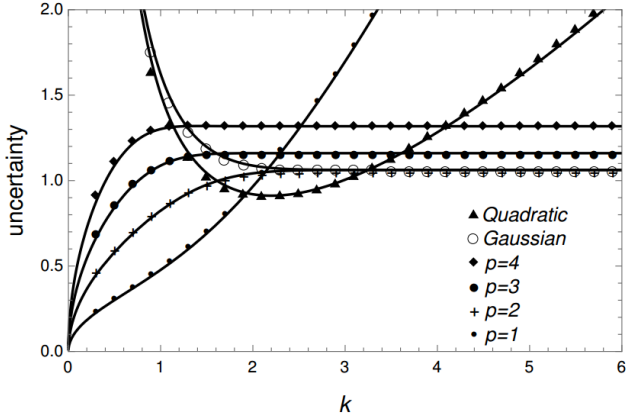

Figure 1. A plot of the uncertainties as a function of $k$ for the generalized centroid for $p = 1, 2, 3$ and 4, the quadratic fit and the Gaussian fit. The curves are from theory and the symbols represent results from simulation. This data is for a one-dimensional Gaussian signal with Gaussian noise assuming a symmetric range centered about the maximum value.

The most noticeable aspect of the curves in Figure 1 is that the uncertainties for the first four powers of the centroid decrease to zero as the range goes to zero. This does not make intuitive sense because as the amount of information decreases, the uncertainty should increase. The reason for this behavior is that we are using the range about the known value of the center. In practice, one first finds the maximum value and assumes this is near the actual peak. Because of noise, the maximum value is not always at the correct value, thus causing an asymmetry in the signal range that we ignored in our derivation. In short, our analysis and the simulations in Figure 1 are missing a crucial step in the peak localization algorithm and this results in a decreasing uncertainty as $k \to 0$.

Another noteworthy feature in Figure 1 is that the quadratic fit has a smaller uncertainty than the Gaussian fit for values of $k \lesssim 3$. But as more data is included from the tails of the Gaussian, the quadratic fit uncertainty becomes increasingly large.

In Figure 2 we present simulations identical to those used in Figure 1, except that we use the maximum value, then acquire a range of data about that value with which to perform computations.

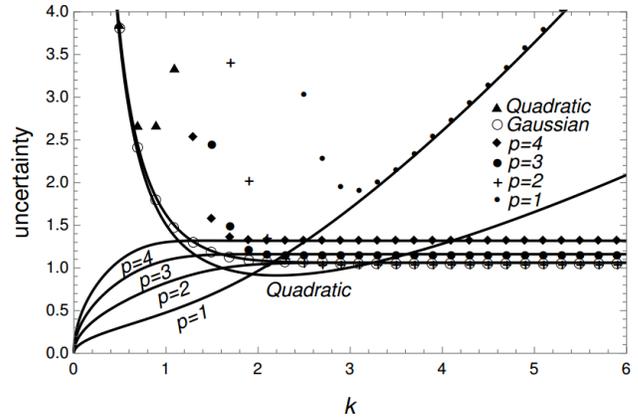

Figure 2. A plot of the uncertainties as a function of $k$ for the generalized centroid for $p = 1, 2, 3$ and 4, the quadratic fit and the Gaussian fit with the maximum value used to determine the center of the range. The curves are from theory and are now labeled because the simulation results do not lie directly on top of them, except for the Gaussian curve.

The simulations in Figure 2 no longer match the theory (except for the Gaussian fitting). These results make more intuitive sense because the generalized centroid uncertainties appear to go to infinity as the range approaches zero.

The centroid methods all quickly approach their theoretical values as $k$ increases. This suggests that small asymmetries in the range about the maximum value become less important as more information is used. We speculate that this behavior is probably only true for signals with a reasonably high signal-to-noise ratio. Note that the noise in our simulation is relatively small, i.e. less than 1/20[th] of the signal amplitude.

The quadratic uncertainty values however, are sensitive to asymmetries in the range. The simulations agree with theory for small values of $k$, but quickly separate from theory and remain far above our theoretical values for most of the range of $k$ tested. In contrast, the Gaussian uncertainty is robust to asymmetries in the range.

We show comparisons for simulations and theory for a 2D Gaussian signal in Figure 3.

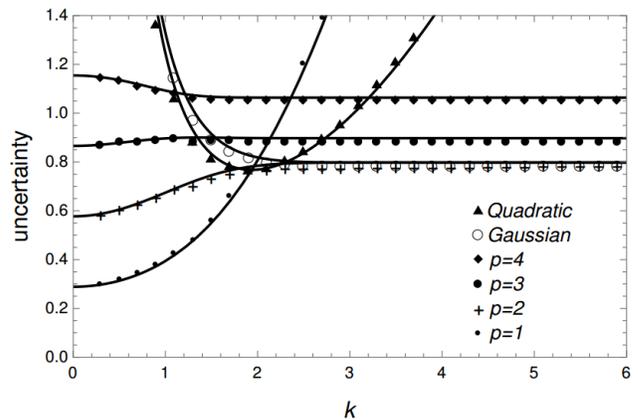

Figure 3. A plot of the simulation results (labeled symbols) and theory (curves) for uncertainties for a 2D Gaussian signal as a function of $k$ for the generalized centroid for $p = 1, 2, 3$ and 4, the quadratic fit and the Gaussian fit with the range of data symmetrically chosen around the center value. The curves from theory are not labeled because the simulation results lie almost directly on top of them.

There is good agreement between theory and simulation. However, as before, the theory assumes that the curve is centered at zero and that the range of data varies symmetrically about the center.

In Figure 4, we use the maximum value as the center and the simulation results again change significantly.

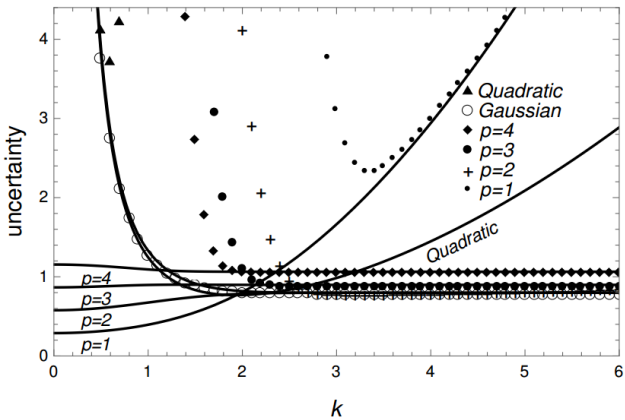

Figure 4. A plot of the peak localization uncertainties as a function of $k$ for the generalized centroid for $p = 1,2,3$ and 4, the quadratic fit and the Gaussian fit with the maximum value used to determine the center of the range. The data are for 2D Gaussian signals. The curves from theory are now labeled because the simulation results do not lie directly on top of them, except for the Gaussian curve.

As in the one-dimensional case, all peak localization uncertainties now increase to infinity as the range decreases to zero. The Gaussian uncertainty is robust to asymmetries in the range. The centroid methods are robust to this asymmetry for large enough $k$. The quadratic uncertainty is again highly sensitive to asymmetry and only a few simulation results are visible at the top of the graph for $k < 1$.

In Figure 5 we display simulation and theory for a one-dimensional Lorentz-like signal with uniform Gaussian noise.

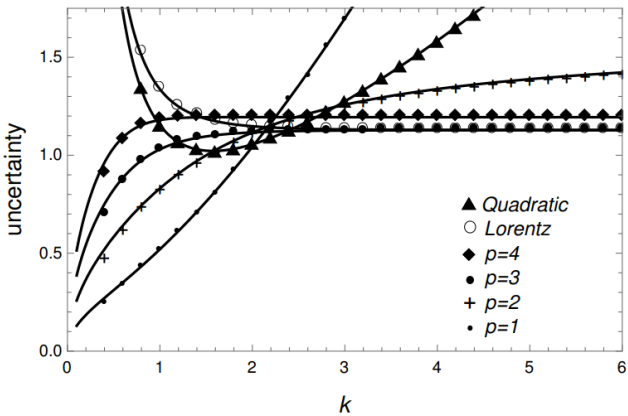

Figure 5. A plot of the uncertainties as a function of $k$ for the generalized centroid for $p = 1,2,3$ and 4, the quadratic fit and the non-linear least squares fit with the Lorentz-like equation. The curves are from theory and the symbols represent results from simulation. This data is for a one-dimensional Lorentz-like signal with Gaussian noise assuming a symmetric range centered about the maximum value.

Our theory and simulations are consistent as shown in Figure 5. Performing a non-linear least-squares fit with the Lorentz-like equation results in the same variance as the centroid of cubes in this case, different from results with the Gaussian signal.

We now include the maximum value in our simulations with the Lorentz-like equation and compare that with our analytical results in Figure 6.

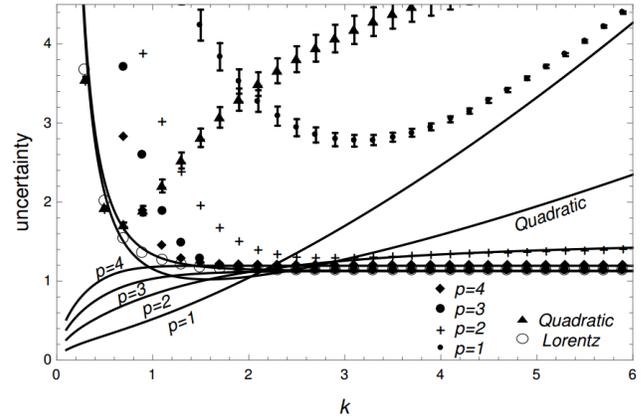

Figure 6. A plot of the uncertainties as a function of $k$ for the generalized centroid for $p = 1,2,3$ and 4, the quadratic fit and the non-linear least squares fit to the Lorentz-like equation with the maximum value used as the center of the range. Estimated 95% confidence intervals are shown for the centroid and the quadratic simulations.

The non-linear fit using the Lorentz-like equation is robust with respect to asymmetries in the range about the center of the curve. The centroid of squares and higher powers are robust to these asymmetries, approaching their theoretical values with values of $k$ from 1-3. However, the centroid approaches its theoretical curve at a slower rate.

The quadratic fit is again the least robust with respect to inclusion of the maximum value.

Note that the confidence intervals become smaller as they approach the analytical curves.

## Summary and Discussion

We compare several peak localization methods both analytically and through simulations for a Gaussian signal and a Lorentz-like signal (both with Gaussian noise). For the Gaussian signal, for large ranges, more than 3 times the width of the curve, i.e. $k > 3$, we find that the centroid of squares and the non-linear least squares fit with a Gaussian function yield the best results. For the Lorentz-like signal, the centroid of cubes and the non-linear least squares fit with the Lorentz-like function give the best results. The non-linear least-squares fits give the best results for small ranges.

Our analysis did not include one crucial aspect of most peak localization algorithms, that of first finding the maximum value and then using a symmetric range about that value as input to the peak localization method. When we include this step in our simulations we find the following:

- The uncertainties approach our analytical results for the generalized centroid for $k > 3$ for $p \geq 2$. The for centroid, the simulations approach our theoretical results more slowly.
- The uncertainties for the non-linear least squares fit to the Gaussian and Lorentz-like signal are unchanged.
- The uncertainties for the quadratic fit are substantially larger than our analytic results for $k > 1$.

Based on our results, we believe that the centroid and the quadratic fit methods should be avoided for peak localization tasks. For windows of data with $k > 3$, the centroid of squares is as good as a non-linear least-squares fit for Gaussian shaped signals. This coupled with its computational speed and ease of implementation indicate that it should be more widely used than it currently appears to be.

For a Lorentz-like curve we find that the centroid of cubes gives identical results to a non-linear least-squares fit for large ranges. Because many peaks in real data are well approximated by a Gaussian or a Lorentzian shape, or some combination thereof, it is interesting to speculate that the centroid of the data raised to some small power, 2 or 3, may be both quick and near optimal in terms of peak localization for a variety of peak shapes.

Future work on this topic should expand the analysis to include the maximum value. Although this complicates the mathematics, it may be feasible for some cases. We should also examine these methods with Poisson noise as well as for more realistic scenarios in which the pixel size is considered.

Lastly, these methods should also be compared when the signal to noise ratio becomes small. This situation is common in application and may change our results.